\documentclass[a4paper,prf,aps,10pt]{revtex4}

\usepackage{amsmath}
\usepackage{amssymb}
\usepackage{graphicx}
\usepackage{fancyhdr}											

\usepackage{hyperref}
\hypersetup{
    colorlinks= true,
    citecolor = blue,
    linkcolor= blue,
    filecolor= magenta,      
    urlcolor= red,
}
\usepackage{epstopdf, epsfig}
\usepackage{color}

\usepackage{textgreek}

\definecolor{orange}{rgb}{0.8500, 0.3250, 0.0980}
\definecolor{amber}{rgb}{1.0, 0.75, 0.0}
\definecolor{arsenic}{rgb}{0.23, 0.27, 0.29}
\definecolor{battleshipgrey}{rgb}{0.52, 0.52, 0.51}
\definecolor{charcoal}{rgb}{0.21, 0.27, 0.31}
\definecolor{darkelectricblue}{rgb}{0.33, 0.41, 0.47}
\definecolor{firebrick}{rgb}{0.7, 0.13, 0.13}
\definecolor{azure}{rgb}{0.0, 0.5, 1.0}
\definecolor{purple}{rgb}{0.63, 0.36, 0.94}

\include{xincludegraphics}

\newcommand{\SImum}{\textrm{\textmu{}m}}

\begin{document}

\title{Dissolution of microdroplets in a sparsely miscible liquid confined by leaky walls}


\author{Jia Ming Zhang}
\email{jiaming.zhang@utwente.nl}
\affiliation{Physics of Fluids group, Max-Planck Center Twente for Complex Fluid Dynamics, Department of Science and Technology, Mesa+Institute, and J. M. Burgers Centre for Fluid Dynamics, University of Twente, Enschede, Netherlands}

\author{Yibo Chen}
\affiliation{Physics of Fluids group, Max-Planck Center Twente for Complex Fluid Dynamics, Department of Science and Technology, Mesa+Institute, and J. M. Burgers Centre for Fluid Dynamics, University of Twente, Enschede, Netherlands}

\author{Detlef Lohse}
\affiliation{Physics of Fluids group, Max-Planck Center Twente for Complex Fluid Dynamics, Department of Science and Technology, Mesa+Institute, and J. M. Burgers Centre for Fluid Dynamics, University of Twente, Enschede, Netherlands}

\author{Alvaro Marin}
\email{a.marin@utwente.nl}
\affiliation{Physics of Fluids group, Max-Planck Center Twente for Complex Fluid Dynamics, Department of Science and Technology, Mesa+Institute, and J. M. Burgers Centre for Fluid Dynamics, University of Twente, Enschede, Netherlands}





\begin{abstract}

When a water droplet is deposited within a sparsely miscible liquid medium such as certain oils, the droplet surprisingly vanishes, even in a confined geometry. Such a phenomenon has crucial consequences for multiphase flows in which confined nano- and/or picoliter droplets are considered. We report here experiments of microdroplet dissolution in microchannels that reveal an enhancement of the shrinkage of confined water microdroplets in oil due to the permeability of the walls - made of polydimethilsiloxane (PDMS) - and a delay when collective effects are present. The system is first modelled assuming that the dissolution of the droplets in its surrounding liquid follows the Epstein-Plesset solution of the diffusion equation. The dissolution of small isolated droplets can indeed be described by this solution of the diffusion equation, while the vanishing of droplets larger than a certain critical value and those closer to other droplets requires numerical simulations which take into account the boundary conditions of the confined system, the neighbouring droplets and interestingly, the evaporative water vapour flux through the PDMS. Our results thus reveal the important role of the water solubility in oil and most remarkably, of the water vapour transport through permeable walls.

\end{abstract}
\maketitle



\section{Introduction}

Diffusive processes in multiphase flows with discrete phases in the nano- or picoliter volume range can yield very surprising phenomena like dissolution or growth of microbubbles \citep{shim2014dissolution,volk:2015}, the spontaneous nucleation of nano-sized droplets in ternary systems \citep{Lohse2015RMP} or, as it will be considered in this work, the vanishing of a droplet in sparsely miscible media. Such systems with well-controlled interfacial properties \citep{stone2004engineering} are found in a large variety of systems as in solvent extraction applications \citep{rydberg2004solvent,jain2011,rezaee2006,rezaee2010,Lohse2015RMP}, emulsion-based DNA sequencing \citep{margulies2005genome}, single-molecule analysis \citep{diehl2006beaming}, designed microemulsions \citep{shah2008designer} or even molecular gastronomy \citep{this2002molecular,this2005molecular}. Such emulsions can become unstable by phase separation induced by coalescence, ripening or sedimentation. 

{Emulsions can be brought out of equilibrium when the discrete phase evaporates (to a neighboring gas phase) or dissolves into the continuous liquid phase.
}
In some systems this is not an issue since the discrete liquid phase is forced into a phase change in order to preserve its content \citep{Takeuchi:2005dl}, or it is directly extracted to be further processed. However, in many other cases, the dispersed phase needs to be stable for longer times, either to allow for mixing within the droplets \citep{Song:2003bg}, to let the solvent evaporate/dissolve or, in the case of bubbles, to allow their gases dissolve in the surrounding liquid \citep{shim2014dissolution}. In many of these cases, the volume loss of the dispersed phase into the host medium can be quite harmful.
Nonetheless, in other cases the dissolution of the discrete liquid phase microdroplets is actually required. This is the case of drying of colloidal suspensions in microchannels \citep{yi2003, wang2017}, microparticle aggregate synthesis \citep{velev2000,manoharan2003,brugarolas2013}, protein crystallization in emulsions \citep{zheng2003,yu2012} or generation of microcapsules \citep{zhang2012one}. In all these examples the rate of dissolution of the dispersed phase into the continuous phase has a crucial role in the way the solute contained in the droplet aggregates, be it with polymeric solutions \citep{spraydrying1998} or with solid particles aggregating in the bulk \citep{Vogel2018magic} or at a droplet's interface \citep{Lauga:2004qy}.

The case of a gas bubble dissolving in an infinite liquid environment at rest was analytically solved in a classical paper by \citet{epstein1950}, which was later extended to liquid droplets immersed in the bulk of a partially miscible liquid phase by \citet{duncan2006}. {There has been considerable recent progress in the understanding of systems of dissolving bubbles or droplets in complex scenarios. In a recent paper, \citet{michelin2018collective} studied theoretically a system of diffusively dissolving microbubbles experiencing collective effects in different distributions in two and three-dimensional geometries, giving a large overview on the different phenomena that can be observed. Three-dimensional collective effects between dissolving bubbles have been recently studied experimentally in micro-gravity conditions and compared with simulations by \citet{Patri2020}. But the phenomenology changes dramatically when convective flow is present, this is the case interacting sessile droplets in a channel immersed in a fluid in motion \citep{Laghezza:2016zr,chong2020convection}. Regarding the confined dissolution of droplets and bubbles, \citet{Rivero2019} have recently studied the steady dissolution of trains of bubbles in a cylindrically-shaped flow, also taking into account the mutual interaction among bubbles. Collective effects such as coarsening and competitive growing have also been shown to take place among nanobubbles \citep{Zhu:2018} in semi-confined conditions.}

Micro-confined conditions can yield a higher degree of control on droplet dissolution \citep{stone2004engineering,zhu2017}. However, the presence of confining boundaries clearly has an impact in the way the dissolution takes place, either to enhance the process or to hinder it. {Polydimethilsiloxane, widely known as PDMS, is the most widely used material for constructing microfluidics chips using soft lithography \citep{xia1998soft}. Before its use in microfluidics, this material had been actually used as membrane for its excellent permeability to vapour and other gases \citep{Rubber1968}. Such feature was soon exploited in early microfluidic research to promote fluid motion in microfluidic channels \citep{DoylePNAS2005}, concentrate colloids \citep{Verneuil2004permeation} or to crystallize salts \citep{LengPRL2006microevaporators}. More recently, PDMS has been used to emulate the vapour transport in leaves \citep{noblin2008optimal,wheeler2008transpiration,dollet2019drying} or to explore evaporation-induced cavitation in droplets inside permeable viscoelastic gels \citep{vincent2012birth,bruning2019turning}.
}
In the present work we take advantage of microfluidic droplet generators to experimentally study the dissolution of both isolated and groups of picoliter droplets in a sparsely miscible medium, confined by rigid but water-permeable walls, as it is customary in microfluidic systems. To do so, we study the shrinkage of water droplets of different sizes in silicone oil inside microfluidic channels made of PDMS. The experimental data are compared with analytical solutions of the Epstein \& Plesset equation \citep{epstein1950}, which assumes an infinite and unconfined external medium. 
This assumption only applies for droplets small enough to ignore the presence of the confining walls. For larger droplets, we proceed to compute numerical solutions of the diffusion equation using an immersed boundary method, which yields results that compare well to the experiments. Remarkably, the evaporative flux of water vapour through the permeable PDMS wall turns out to be crucial to obtain good agreement with the experimental data. This point is key for microfluidic long-term processes carried out in PDMS-based devices.

The paper is organized as follows: First, we describe the experimental setup and results in section \ref{sec:experimental}. In section \ref{sec:modeling} we compare the experimental results with those of the Epstein-Plesset equation. It will be shown that the analytical solution of the Epstein-Plesset equation is insufficient to describe all experimental results. Therefore numerical solutions are required, which will be shown and discussed in section \ref{sec:numresults}. The paper ends with an outlook and conclusions in section \ref{sec:conclusions}.

\begin{figure}
  \centerline{\includegraphics[width=0.75\linewidth]{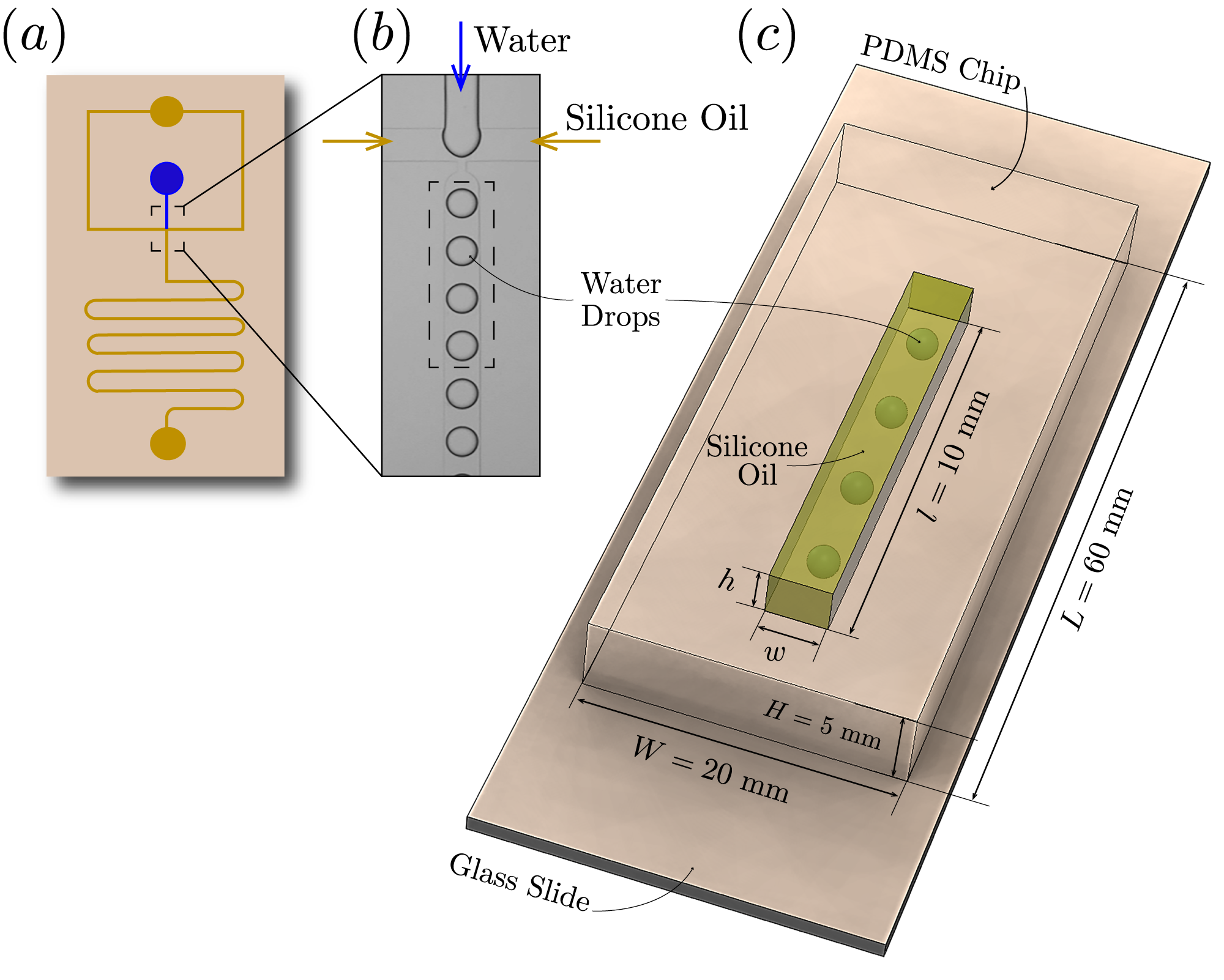}}
  \caption{\textcolor{black}{Schematics of the setup employed to study water microdroplet shrinkage within silicone oil in microchannels: (a) Microfluidic chip design (b) Close-up view of the droplet generation  junction (flow focusing). (c) View of the PDMS chip and glass slide. The inner dimensions of the main channel used in the experiments are $h =$ 85 $\SImum$, \textcolor{black}{$w =$ \textcolor{red}{40}--104 $\SImum$}, and $l =$ 10 mm.}}
\label{fig:schematic}
\end{figure}

\section{Experiments}\label{sec:experimental}

\subsection{Experimental setup}

Figure \ref{fig:schematic} shows sketches of our PDMS microfluidic device to study the dissolution of microdroplets. The sketches' purpose is only illustrative and therefore the proportions are not realistic in the figure. The microfluidic chip was fabricated using soft-lithography techniques \citep{xia1998}, with a mixing ratio of Sylgard 184 base and curing agent of 10:1. A thin film of PDMS was also spin-coated on the bottom glass slide to ensure the hydrophobicity of the channel. Figure \ref{fig:schematic}(a) shows the chip design and Figure \ref{fig:schematic}b shows a close-up of the microfluidic flow-focusing structure: The dispersed phase (deionized water) is forced downstream the 4-way junction through the constriction by the continuous phase (20 cSt silicone oil, Sigma-Aldrich). As a consequence, the water meniscus breaks into highly monodisperse water microdroplets (Figure \ref{fig:schematic}b). Once the downstream serpentine channel is filled with the dispersed phase, the flow is completely stopped and the shrinkage measurements begins. By tuning the flow rate ratio of the two phases, different sizes of microdroplets and spacing between microdroplets can be achieved. In this case, we worked with microdroplets with radii in the range \textcolor{black}{$4\;\SImum - 40\;\SImum$}, covering three orders of magnitude in volume: from picoliter up to nanoliter droplets. Two three-way valves were used in the water and oil supply path, respectively. For generating microdroplets, two syringe pumps (Harvard high-precision syringe pumps) supplied fluids through the valves into the chip. When the desired number of microdroplets were generated, the three-way valves were switched to cut-off the fluid supply and immediately opened to expose the main channel to atmospheric pressure. In such a way, the pressure in all inlets was swiftly switched to atmospheric pressure and all pressure gradients within the channel quickly died out. Following this procedure, a number of microdroplets can be fully stopped in the serpentine channel for dissolution experiments.
It is important to note that, although surfactants are typically used in these systems to stabilize the emulsion and prevent droplet coalescence, the use of the tiniest amount of surfactants can have crucial consequences in the  process. Since we want to focus on the dissolution/shrinkage process only, no surfactants were used in any of our experiments.

The microfluidic chip was placed on an inverted optical microscope using $10\times$ and $60\times$ objectives (Nikon). A CCD camera (Ximea) was used to record the  microdroplets shrinkage. \textcolor{black}{The optical system yielded a final resolution of 0.12 $\SImum$/pixel.} Image analysis and measurements were performed using home-made Matlab codes. All experiments were conducted at room temperature of 22 $^{\circ}$C and \textcolor{black}{between 30\% and 40\% relative humidity}.

\subsection{Experimental conditions and assumptions}

Experimental conditions need to be defined in order to understand how the experiments were performed. Both the silicone oil and the PDMS chip were degassed prior to the experiments to ensure that no water vapour was pre-absorbed prior to the experiments. We will therefore assume that the initial vapour concentration in oil and within the PDMS walls is zero.

{Given the density difference between the two phases involved, we should consider the possibility of buoyancy effects {within the continuous phase} during the droplet shrinkage.
This can be evaluated by computing the ratio between the viscous time scale and the convective time scale induced by buoyancy, i.e. the Grashof number $Gr={g\Delta\rho R^3}/{(\rho\nu^2 )}$, with $g$ the gravitational acceleration, $\Delta \rho$ is the maximum density difference between the liquid phases, $\rho$ is the density of the continuous medium ($\rho$ = 950 $kg/m^3$) and $\nu$ its kinematic viscosity (20 cSt). In our case $Gr$ takes values in the range from $10^{-7}$ for the smallest droplets to $10^{-4}$ for the largest, and therefore we will neglect buoyancy effects in the main discussion of the paper.}

\begin{table}
  \begin{center}
  \begin{tabular}{|c|c|c|c|}
     Confinement  & $c=R_{0}/R_{B}$ & $R_{0}\,(\SImum)$  & $R_{B}\,(\SImum)$ \\[2pt]  \cline{1-4} \\[-9pt] 
       Weak   & $0.1 - 0.3$ & $4 - 8$ & $20 - 45$ \\       
       Strong  & $0.3 - 0.8$ & $20 - 40$ & $40 - 52$\\ 
  \end{tabular}
  \caption{Classification of experiments according to different initial microdroplet sizes and channel dimensions, casted into the confinement ratio $c$, defined as the ratio between the initial droplet radius $R_0$ and half channel's smallest dimension (typically width) $R_B$.}
  \label{t1}
  \end{center}
\end{table}

The experimental results are organized according to the degree of confinement of the droplets in the channel. In table \ref{t1} we summarize three series of experiments performed, classified by the confinement ratio which is defined as $c=R_{0}/R_{B}$, where $R_{B}$ is the half width of the channel and $R_{0}$ is the initial microdroplet radius. Note that the droplet is fully surrounded by silicone oil and there is no contact between the microdroplet and any of the channel walls. Both height and width of the channel are larger than the initial microdroplet size, so only spherically-shaped droplets are considered in this study. Consequently, only droplets with (projected) diameter smaller than the channel height ($h=$ 85 \SImum) were considered. {The distinction between weak and strong confinement is chosen according to the amount of water that can be dissolved by the oil phase in our geometry.  Considering an isolated water droplet (density $\rho_w$) in a channel of length $10^4 \, \SImum$, {width $2R_B=100 ~\SImum$} and depth $85 \,\SImum$ filled with oil ($V_{oil}$), and the saturation density of water in oil ($ \rho^{o}_{w} =0.2 \, \mathrm{kg/m^{3}}$, taken from \cite{garbay1984}, see discussion below), we obtain that the maximum amount of water that can be dissolved by such volume of silicone oil is $V_{c} = V_{oil}\times \rho^{o}_{w}/\rho_{w}=17$ pL, which corresponds to a water droplet of radius $R_c= 16~\SImum$. We can define then a critical confinement ratio $R_c/R_B = 0.32$. Consequently, all droplets with $R_0<R_c$ ($c<0.3$) are considered as \emph{weakly-confined} and those with $R_0>R_c$($c>0.3$) as \emph{strongly-confined} (Table \ref{t1}).}

{A comment deserves to be done regarding the position of the droplets within the channel. Due to the density difference between both liquid phases, an initially central-positioned droplet would in principle experience some displacement due to gravity while it dissolves, depending on its size. A simple calculation can show that droplets smaller than about 7 $\SImum$ are displaced a negligible amount during their lifetime. But larger droplets would experience larger displacements. However, since they are typically confined, the distance displaced is very much limited by the channel wall itself and the effect is consequently constrained. Therefore, we do not expect that displacements from the center of the channel have a substantial effect on the dissolution process.}

\begin{figure}
  \centering
  \includegraphics[width=0.9\textwidth]{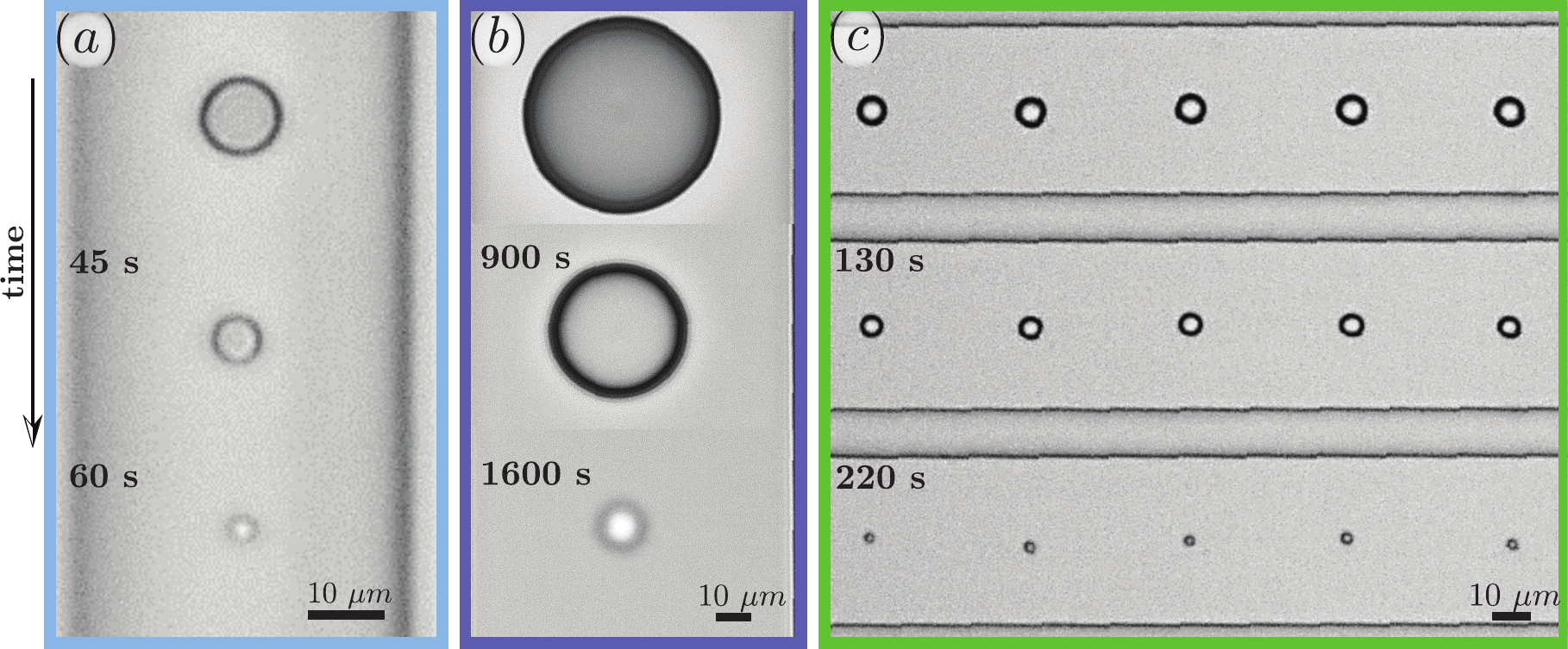}
  \caption{Sequence of snapshots of the shrinkage process of microdroplets in a microchannel (a) isolated under low confinement ($R_0=$ 5 $\SImum$, $c=0.1$), (b) isolated under high confinement ($R_0=$ 31.5 $\SImum$, $c=0.60$) and (c) group of droplets under low confinement ($R_0=$ 3.5 $\SImum$, $c=0.17$).}
  \label{fig:sequence_drops}
\end{figure}

\subsection{Experimental results} \label{sec:allexp}

A sequence of images showing typical shrinkage experiments can be found in Figure \ref{fig:sequence_drops}, corresponding to (a) a droplet in \emph{weak confinement} and (b) a droplet in \emph{strong confinement}. {We define a droplet as \emph{isolated} when the closest droplet is at a distance of at least 15 diameters. The origin of this particular reference value for the droplet distance will be developed in section \ref{sec:numresults}.} However, since droplets are not generated individually and in isolation in typical operating situations with microfluidic devices,  
Figure \ref{fig:sequence_drops}(c) also show a group of droplets in \emph{weak confinement}. They are instead found in equally spaced groups of droplets \citep{wang2017,yi2003,shim2014dissolution,Takeuchi:2005dl,shah2008designer}. In a typical experiment, hundreds or thousands of droplets are generated per second and they fill the whole device. Unfortunately, droplet size and spacing are strongly correlated, which does not allow us to modify such parameters independently. The case shown in Figure \ref{fig:sequence_drops}c shows the dissolution of a group under weak confinement ($c=0.17$) with an almost constant spacing ($5\times$droplet size), in which they all remarkably shrink at the same pace.

\begin{figure}
  \centering
  \includegraphics[width=0.55\linewidth]{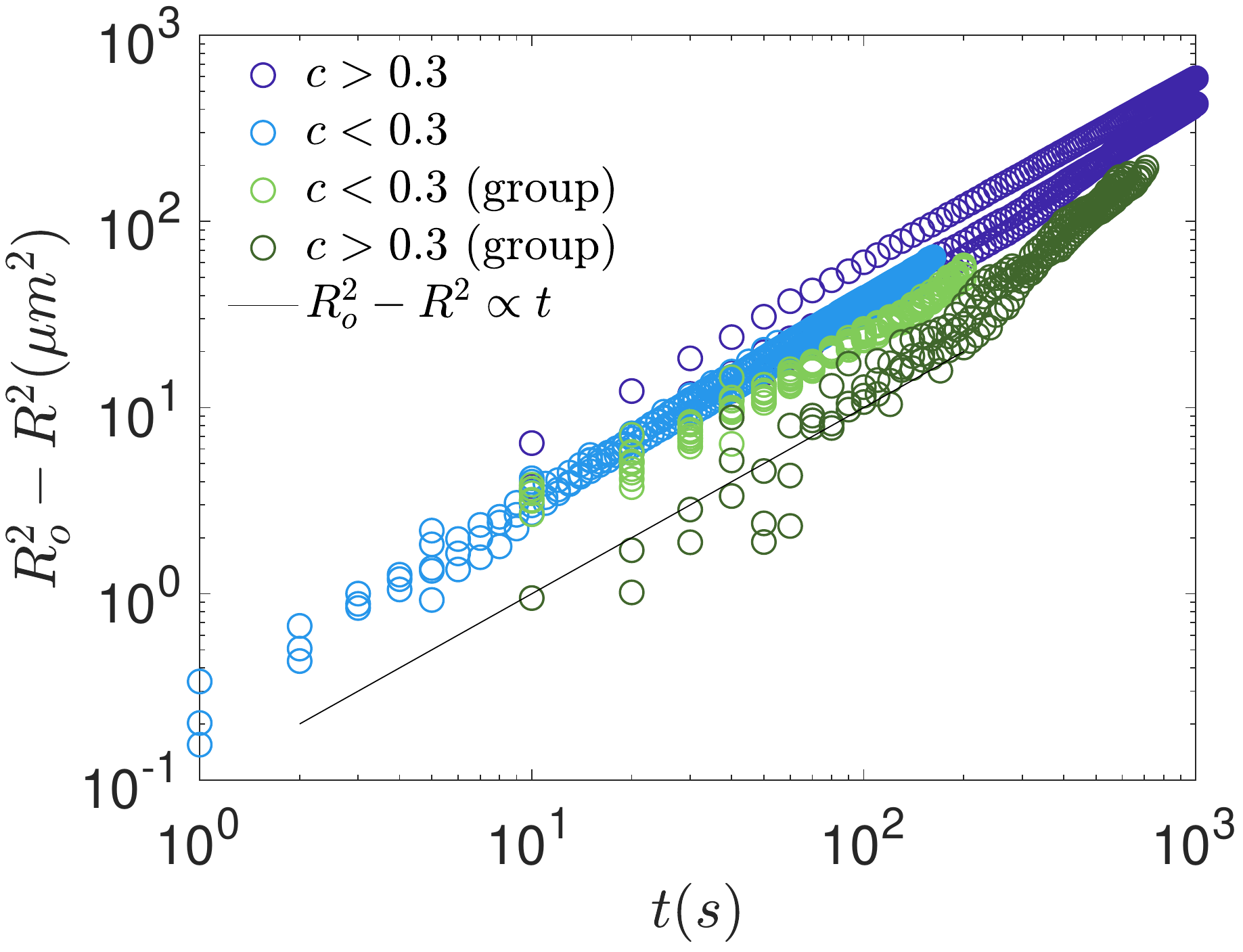}
  
  \caption{Data from all experiments ran for this study, plotted as $R_o^2-R^2$ against time, including single droplets dissolving in weak-confinement (i.e. their initial diameter is below 30\% of the total channel width $c<0.3$), single droplets in strong confinement $c>0.3$, and group droplets in both situations. The data is compared with the linear law $R_o^2 -R^2 \propto t$, typically found in diffusive processes.}

\label{fig:alldrops_plot}
\end{figure}

Figure \ref{fig:alldrops_plot} shows measurements in a wide range of droplet radii and conditions (weakly and strongly confined, single and grouped). Dissolution times range from a few dozens of seconds in the case of single weakly-confined ones ($c<0.3$, as in Fig. \ref{fig:sequence_drops}a), to more than 1000 s for the single strongly-confined droplets ($c>0.3$,  as in Fig. \ref{fig:sequence_drops}b). As can be seen in Figure \ref{fig:alldrops_plot}, most droplets vanish following $R_o^2-R^2 \propto t$, from which we can infer that the process is entirely diffusive. Consequently, in the following section we will employ the Epstein-Plesset equation to capture the diffusive shrinkage of single weakly-confined droplets and compare the results with the experimental data.  



\section{Epstein-Plesset model for droplet dissolution}\label{sec:modeling}

The Epstein-Plesset equation (EP)\citep{epstein1950} was originally developed to describe the dissolution of a single and isolated spherical gas bubble in an infinite liquid medium and it was later successfully applied to describe also the dissolution of droplets immersed in the bulk of partially miscible liquids by \citet{duncan2004} (for more details and recent developments, as the extension to sessile droplets and bubbles, see \cite{Lohse2015RMP}). 

In the geometry considered here, the concentration of the dispersed phase in the far-field liquid-phase $C_{\infty}$ is considered to be undersaturated at $t=t_0$ ($C_{\infty}<C_S$), and the concentration of the dispersed liquid at the droplet surface is considered to be at saturation $C_{S}$. Unfortunately, the data on the saturation concentration of water in silicone oil in the literature are scarce. 
{\citet{garbay1984} obtained the  amount of water absorbed by silicone oil at different humidities. Since we are interested in the maximum amount of water in the close vicinity of the droplet in the oil phase, we use their value obtained at 100\% relative humidity, and identify this value with the saturation concentration of water in oil $C_{S}=0.2 ~\, kg/m^{3}$.}
Regarding the diffusion coefficient $D_\textrm{o}$ of water in silicone oil, some studies in the literature \citep{hilder1971,cussler2009} have reported values of water diffusivity in various different solvents, with values in the order of $10^{-9}~m^2/s$. However, since the conditions of such experiments are not identical as ours, we choose to use the water diffusivity in oil $D_\textrm{o}$ as the single fitting parameter in our study, keeping the values reported in the literature for other solvents as a reference. 
Under these conditions, the steady-state dissolution rate $dR/dt$ of a liquid droplet can be approximately expressed as

\begin{equation}
  \frac{dR}{dt}=-\frac{D_\textrm{o}(C_{S}-C_{\infty})}{\rho}\frac{1}{R},
  \label{eq:EPS}
\end{equation}

\noindent where $R$ is the droplet radius and $\rho$ is the density of the dispersed phase. Unlike the case of \citet{duncan2004}, steady-state is reached in our process since the typical dissolution time $t$ is much larger than the diffusive time scale, i.e. $t\gg R^2\rho/(\Delta C D_\mathrm{o})$ \citep{Lohse2015RMP}, where $\Delta C =C_s-C_\infty$. The solution of equation (\ref{eq:EPS}) can be written as:
\begin{equation}
\left(\frac{R}{R_o}\right)^2=1-\frac{t}{t_f}, \label{eq:R}
\end{equation}
where $t_f$ is a good approximation for the droplet lifetime that takes the form $t_f=\rho R_o^2/(2D_\mathrm{o} \Delta C)$. The Laplacian pressure slightly increases the dissolution rate for small-sized droplets. However, the typical droplet size at which the Laplace pressure becomes relevant is in the order of $R \sim 1~\SImum$, which is much smaller that the droplet size range covered in our study.

\begin{figure}

\includegraphics[width=\textwidth]{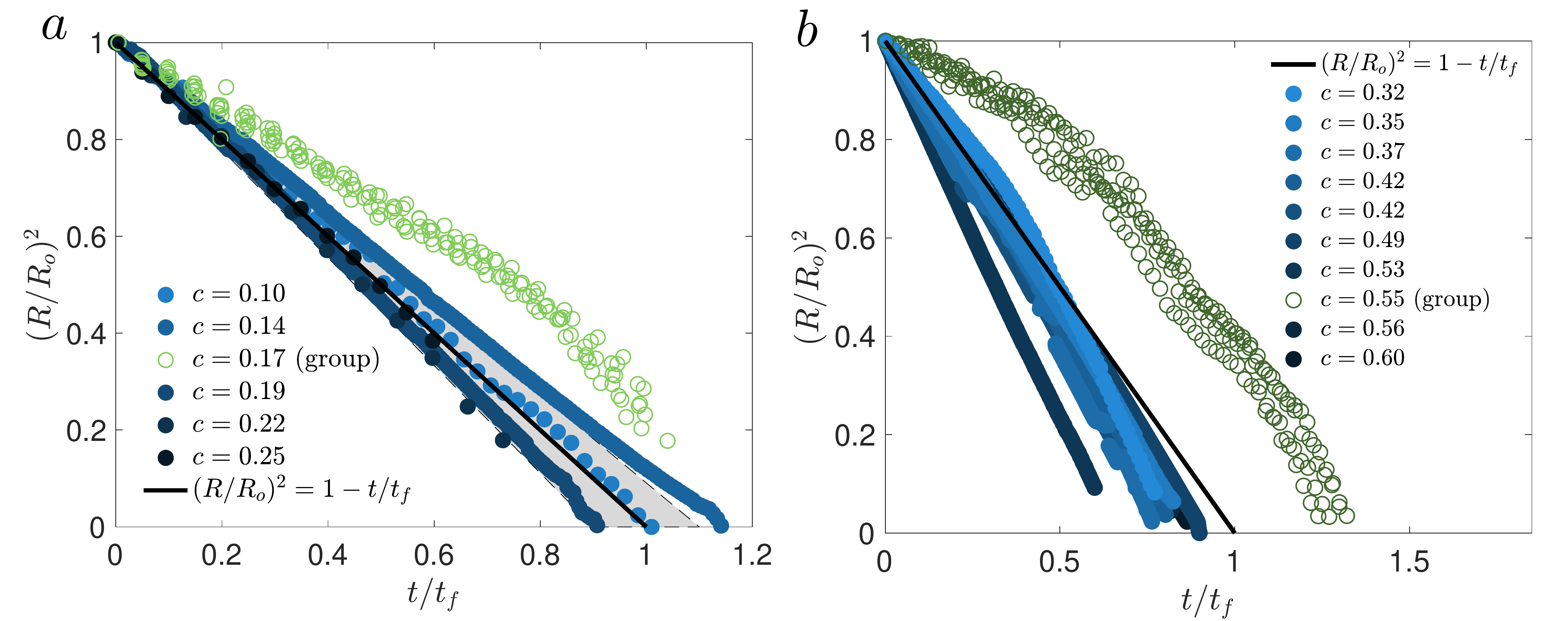}

\caption{Experimental data of dissolution of droplets with different initial radius and different weak confinements. The data is presented in dimensionless form. The continuous black line in both plots corresponds to the analytical solution of equation (\ref{eq:EPS}), i.e. the EP model.
(a) Experiments with droplets under weak confinement $c<0.3$, including group droplets. The grey area surrounding the EP line corresponds to errors of 10\% in the determination of the diffusion constant $D_\mathrm{o}$. (b) Experiments with droplets under strong confinement $c>0.3$.} 
\label{fig:EPSplots}
\end{figure}

Figure \ref{fig:EPSplots}a shows experimental data of the dissolution of droplets in low confinement conditions. The results clearly show that all experiments with single droplets under weak confinement follow very closely the analytical solution of the EP model in equation (\ref{eq:R}). 
Using the data of the weakly-confined droplets, we obtain the best-fit value for $D_\textrm{o}$, namely $D_\textrm{o}=0.94\times10^{-9} \, m^2/s$. Figure \ref{fig:EPSplots}a also shows how the theoretical prediction varies when the diffusivity value changes within a 10\% margin, which is a reasonable precision for the value of $D_\textrm{o}$ found. Measurements of liquid-liquid diffusivities are non-trivial, specially for sparsely miscible liquids as those we are dealing with. Nonetheless, well-controlled dissolution processed based on microfluidics provide a very reliable environment in which the EP equation can be used for obtaining liquid-liquid diffusivities. Small volume droplets (in the range of \textit{pL}) can be fully dissolved in another sparsely miscible liquid and its final dissolution time can be measured accurately. Actually, the reliability of the measurements suggest that, if the solubility of the phases is known a-priori, this could be a robust method to measure the liquid-liquid diffusivity, especially for such a low degree of miscibility among the two phases in which other methods usually fail. 

Also strongly confined droplets are shown in Figure \ref{fig:EPSplots}b. As can be seen, the EP model systematically underestimates the dissolution rate for those droplets under strong confinement, which are the cases that deviate the most from the EP model. Since strongly confined droplets are closer to the PDMS walls, this result clearly shows the important role of the wall permeability to water in the process. 

The results so far concerned single and isolated droplets. We now turn our attention to those droplets found in groups. Droplets laying close to each other are also expected to influence each other's vanishing process. The data in Figure \ref{fig:EPSplots}a also shows grouped droplets in low confinement. This is the same case shown in the image sequence in Figure \ref{fig:sequence_drops}, which shows a remarkably homogeneous shrinkage rate for all droplets, with all vanishing practically at the same rate.
Despite the homogeneity in the group of weakly-confined droplets, the results shown in Figure \ref{fig:EPSplots}(both a and b) deviate strongly from the EP model, which overestimates their diffusion rate, regardless of their degree of confinement. In contrast, as discussed above, the EP model underestimates the dissolution rate of highly confined single droplets. 

Both results, the enhanced dissolution by confinement and the delayed dissolution due to collective effects, evidence the need of a more detailed model in which more realistic boundary conditions can be applied. Indeed, the disagreement of the EP model with the cases of (1) strongly-confined droplets and (2) grouped droplets clearly shows that the hypothesis of dissolution in an infinite medium fails since (1) the proximity of the channel walls has a significant influence on the enhancement of the dissolution and since (2) neighbouring droplets clearly influence each other delaying their overall dissolution. To account for the shrinkage of droplets in confinement, we need numerical solutions of the diffusion equation that account for droplet interactions and for water flux across boundaries.

\section{Numerical solution of the diffusion equation for confined and grouped droplets}
\label{sec:numresults}

Numerical solutions are obtained using an Immersed Boundary Method (IBM). More detailed information about the method and implementation can be found in \cite{Zhu:2018} and in \cite{chong2020convection}. 
{We will obtain numerical solutions of the diffusion equation for the water/vapor concentration $C$ in the different environments surrounding the droplet: oil and PDMS, with coefficients of diffusion $D_o$ and $D_\mathrm{PDMS}$:
\begin{equation}\label{eq:diffusion}
 \frac{\partial C}{\partial t}= D_{\alpha}\nabla ^{2}C.
 \end{equation}
Where $\alpha$ refers to either the oil phase or the PDMS phase. At the droplet's surface $r=R$, the concentration can be assumed to be saturated:
\begin{equation}\label{eq:CS}
C (r=R) = C_S. 
\end{equation}
The droplet is surrounded by 4 walls, where the bottom wall is a glass substrate, impermeable to water/vapor, such that a no-flux condition needs is satisfied: 
\begin{equation}\label{eq:noflux}
\frac{\partial C}{\partial \textbf{n}}=0,
\end{equation}}

For the case of the dissolution of a single droplet, the channel length considered in the simulation along the channel's axis is chosen as 20 droplet diameters, which ensures that the far field boundary condition is satisfied. The channel width is varied depending on the confinement ratio and the channel height for all simulations is $h=85~\SImum$, mimicking the experimental conditions. The grid size for the strong confinement case is set as 1 $\SImum$ and 1/3 $\SImum$ for weak confinement (since these are typically smaller droplets).

In addition to the dissolution in silicone oil, and in order to impose realistic boundary conditions at the channel wall, we need to solve the water vapour transport through the PDMS network. 
{The experiments were performed either with ``freshly baked'' PDMS chips, degassed or recycled ones, through which we forced dry air prior to the experiments. Consequently, in every experiment the PDMS slab is initially completely dry and therefore the initial condition for vapour concentration within the slab can be safely taken as $C_\mathrm{\tiny{PDMS}}$=0 at $t$=0. For simplicity, we will consider that the external finite humidity has no influence in the process (the experiments are run under typical relative humidity between 30\% and 40\%, which corresponds to vapour concentration in the air of $C_\mathrm{air}\le$ 0.007 kg$/\mathrm{m}^3$), and therefore in our numerical model we will consider the PDMS slab as a vapour sink, keeping $C_\mathrm{PDMS}$=0 at $y=L_{\scriptstyle \mathrm{PDMS}}$. Note that using a finite value of humidity at the end of the PDMS slab also requires a detailed numerical model of the PDMS device, which is quite computationally demanding as we will discuss below.}

\begin{figure}
	\centering
	\includegraphics[width=0.60\textwidth]{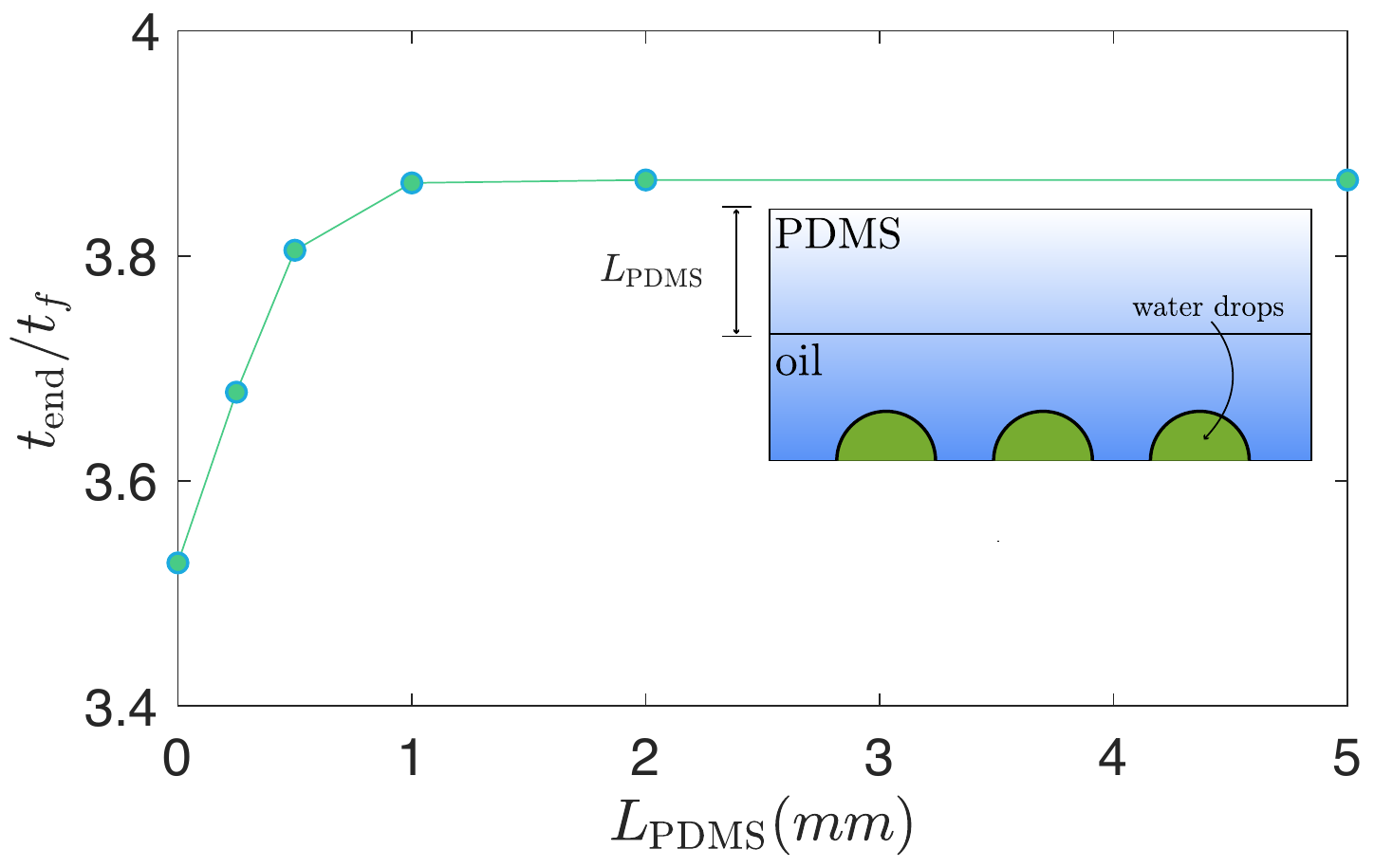}
	\caption{Influence of the wall thickness $L_\mathrm{PDMS}$ on the total dissolution time in the IBM simulations. These simulations are performed assuming a  cylindrical PDMS slab to reduce computational time. The plot shows that the total shrinkage time for a group of weakly confined droplets ($c=0.17$) at different PDMS thicknesses $L_\mathrm{PDMS}$ hardly changes beyond 1 mm.
  The shrinkage time is normalized using the dissolution time $t_f$ expected by the EP equation (\ref{eq:EPS}). The color code in the sketch shown in the inset represents the water concentration in the oil phase (lower part) and in the PDMS (upper part).}\label{fig:Lvst}
\end{figure}

Numerically solving the diffusion of vapour through a fully realistic model of the chip requires high computational costs. Obviously, it would be desirable to reduce its size in the numerical model. In order to test the dependency on the slab size, we have performed tests with a cylindrical PDMS slab geometry, with walls of different thickness to confirm that, when the PDMS wall reaches certain thickness, the total diffusion time does not change significantly. These results are shown in Figure \ref{fig:Lvst}, in which one can see that the error made by considering a PDMS thickness above 500 $\SImum$ is small (below 3\%) in terms of the total diffusion time. {Consequently, the thickness $L_\mathrm{PDMS}$ is set at 500 $\SImum$ for the simulations with rectangular cross-section}. 

\begin{figure}
	\centering
	\includegraphics[width=0.75\textwidth]{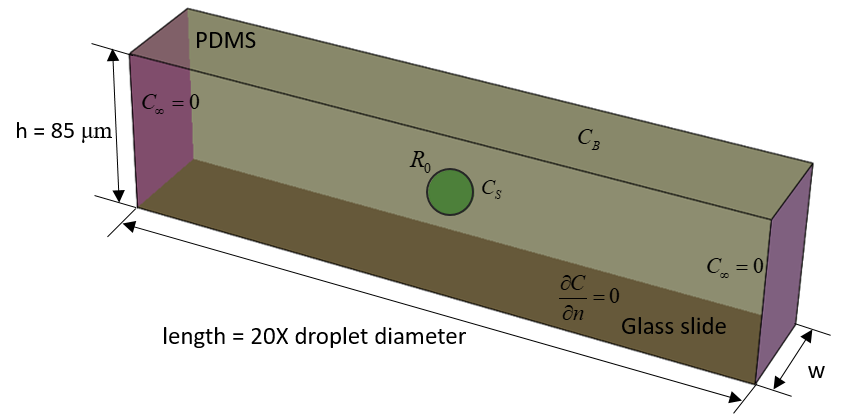}
	\caption{Schematic of the numerical domain and boundary conditions for a single droplet.}
	\label{fig:NumScheme}
\end{figure}

Additionally, although the diffusion constant $D_\textrm{p}$ of water in polymeric materials similar to PDMS has been reported in the literature \citep{watson1996,blume1991,deJong:2006jt}, the conditions, procedures, materials and the proportions of curing agents might vary. Consequently, we choose to take $D_\textrm{p}$ as a fitting parameter, which yields a best fit value of $D_\textrm{p} = 2\times10^{-9}~m^2/s$, similar to values in the literature for similar materials \citep{watson1996,blume1991,deJong:2006jt}. {Interestingly, this value also agrees with those reported in PDMS-based ``microevaporators'', which makes use of the permeation of water through PDMS to induce liquid flows \citep{DoylePNAS2005}, concentrate colloids \citep{Verneuil2004permeation} or to crystallize salts \citep{LengPRL2006microevaporators}. Since the liquid flow rate obtained in these systems depends linearly on $D_{p}$, one can obtain an indirect measurement of the diffusion coefficient of water in PDMS by simply measuring the liquid flow.}

{The similar value of the diffusion coefficient for water in PDMS and silicone could be exploited to simplify numerical simulations and remove the PDMS/oil interface by a single medium with an effective diffusion constant. This similarity in the coefficient values might be an additional reason why the relative position of the droplets respect to the PDMS wall does not seem to play an important role.} 

\begin{figure}
\centering
\includegraphics[trim=0 20 0 0,width=0.55\textwidth]{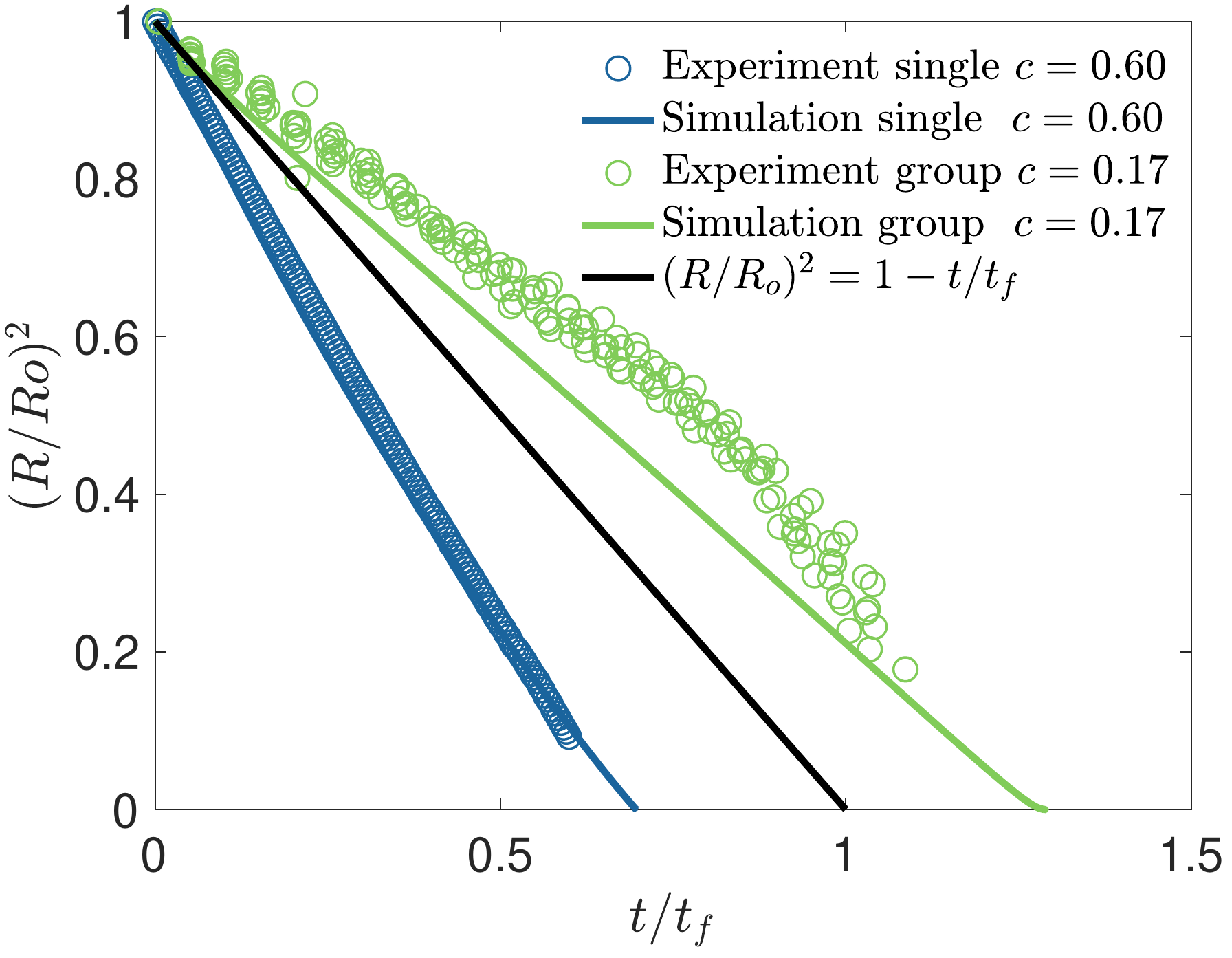}
\caption{Comparison of experimental with numerical results. Single droplet under strong confinement with $R_0 = 22.6 ~ \SImum$ and $c = 0.6$ and group droplets with low confinement $R_o=8.5 ~\SImum$ and $c=0.17$. Continuous lines correspond to simulations and discrete points to experimental data. The black continuous line corresponds to equation (\ref{eq:R}), i.e. the analytical solution of the Epstein-Plesset equation (\ref{eq:EPS}).}
\label{fig:num_results}
\end{figure}

The comparison of the numerical results with the experimental data is shown in Fig. \ref{fig:num_results}. For simplicity, we have chosen one typical case of a single droplet under very strong confinement ($c=0.6$) and a typical case of homogeneous dissolution of group droplets under weak confinement ($c=0.17$). In the case of a single strongly-confined droplet, we can see an excellent agreement of the experimental data with the numerical results with only some minor deviations in the last instants of the shrinkage process. 

Grouped droplets are modelled assuming periodic boundary conditions with no-flux condition ($\partial C/\partial n = 0$) at the mid-plane separating each pair. Such a numerical model assumes that all droplets in a group will shrink at the same rate, which is an approximation consistent with the experimental observations since the droplets in a group have shown negligible differences in shrinkage rate (see Figure \ref{fig:sequence_drops}c). The results of the analytical solution and the numerical one are shown in Figure \ref{fig:num_results} for a group of droplets with an initial weak confinement ratio $c=0.17$. Although the numerical results approach the experimental data closer than the analytical EP model, there is a systematic overestimation of the dissolution rate in the intermediate times which we do not capture with the numerical solution. This deviation has been observed systematically for all experiments with group of droplets and unfortunately we do not have a clear explanation for it. Interestingly, the experimental curve turns towards the numerical one at the later stages of the process. {Our main hypothesis to explain this disagreement is that the precise geometry of the PDMS device, which is not captured by our numerical model, becomes more relevant in grouped droplets than for isolated ones.} In any case, the numerical solution compares significantly better with experiments than equation (\ref{eq:R}) and similar results have been obtained for a wide range of droplet separations and sizes. Unfortunately, droplet size and separation are strongly correlated in microfluidic flow focusing devices and therefore a systematic experimental study on this effect is not straightforward. Nonetheless, this result shows the crucial importance of vapour transport through the PDMS when a significant number of droplets are dissolved simultaneously in a vapour-leaky channel. Note that the vanishing of groups of droplets in a channel with a non-permeable wall would be limited solely to the amount of water capable of being dissolved in the oil phase, which can even lead to an equilibrium with finite droplet sizes at $t\rightarrow \infty$, as $\Delta C \rightarrow 0$ for full saturation and $t_f \propto 1/\Delta C$. 

\begin{figure}
	\centering
	\includegraphics[width=0.65\textwidth]{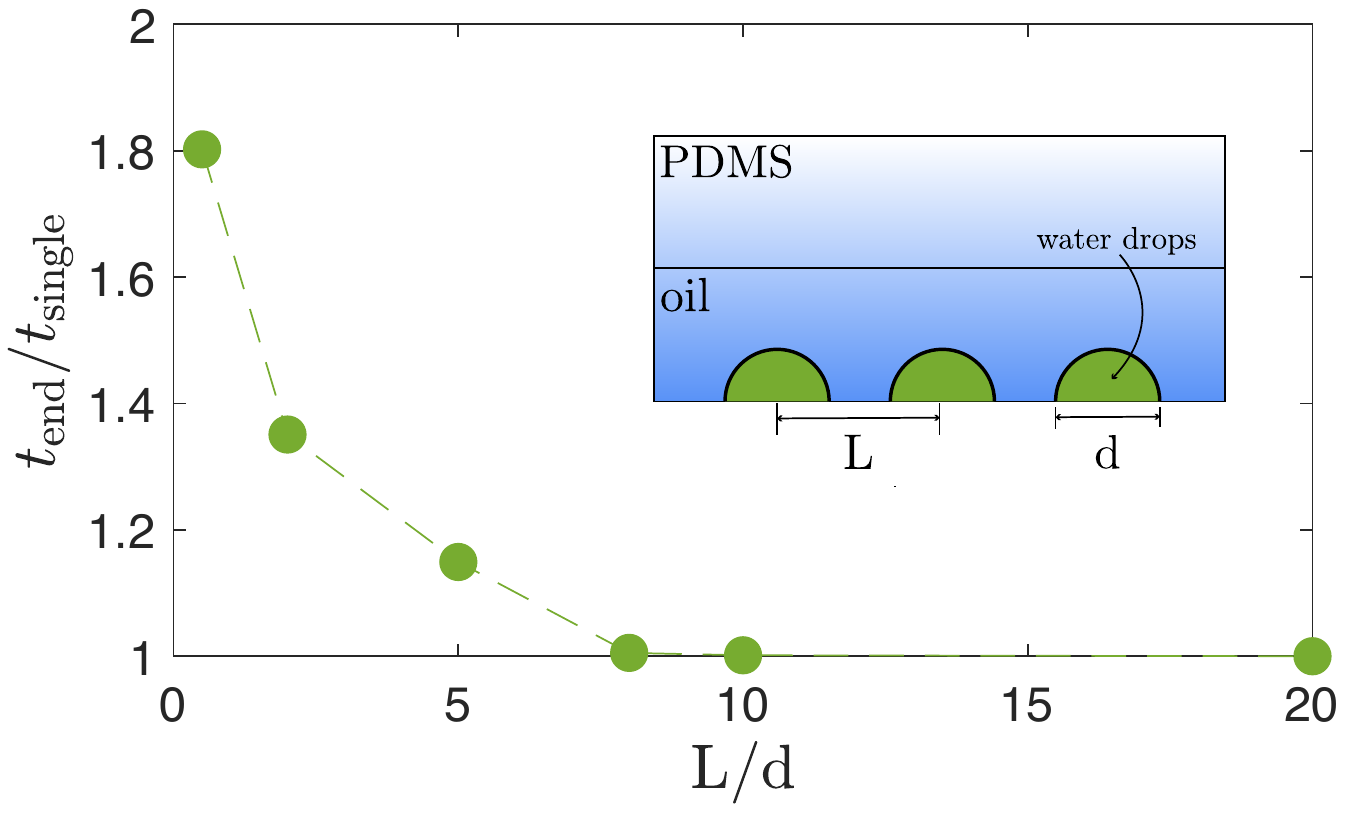}
	\caption{Total dissolution time $t_\mathrm{end}$ for a group of droplets of initial size d, separated by a distance L. Time is normalized by the total life-time of a single droplet dissolving under the same conditions. }
	\label{fig:dropsep}
\end{figure}

Groups of dissolving droplets can show a variety of collective effects. In our case, the droplet dissolution is slowed down significantly due to the presence of neighbouring droplets. Our results raise a natural question: how close do they need to be to show such collective effect? To answer this question, we consider a group of weakly confined droplets, which would follow the analytical unconfined EP model when they are isolated. The initial droplet size chosen is $d=2$ $\SImum$, confined in a cylindrically-shaped channel of diameter 86 $\SImum$ (same as the width of the channel used in experiments). The distance between the droplets is varied from one diameter ($L/d=1$) up to 20 diameters ($L/d=20$), and the total diffusion time $t_\mathrm{end}$ is normalized by the time taken by a single droplet to completely dissolve under the same conditions. 
The normalized shrinkage time for different droplet separation lengths -- normalized by the droplet size -- are shown in Figure \ref{fig:dropsep}. As can be seen, the screening effect of the neighbouring droplets can be ignored when the drop spacing is about 10 times the drop size. Note that without the presence of a permeable wall, the water vapour emanated from the droplets would concentrate within the continuous liquid phase and the collective effects would greatly delay the droplet dissolution.

\section{Conclusions and Outlook}\label{sec:conclusions}

In the present study, a PDMS-based microfluidic device enables us to systematically study the shrinkage of single and multiple droplets of volumes ranging from picoliter to nanoliter, in a confined and sparsely miscible liquid medium. Our results show that the shrinkage occurs via a dissolution of water in the sparsely miscible oil phase, which is then diffusively transported through the water-permeable PDMS. Consequently, the process strongly depends on how much the droplets are confined within the channel. In the weak confinement case ($c<0.3$), the presence of the walls has little effect and the droplet dissolves completely within the oil phase, with no influence from the PDMS matrix. In the strong confinement case ($c>0.3$), the presence of the walls cannot be neglected. To account for the complex geometry, the diffusion equation is solved using Immersed Boundary Method, which yields a better prediction of the faster dissolution of droplets in strong confinement, due to the vapour leak through the permeable wall. Finally, our numerical results reveal an expected slower dissolution for grouped droplets, but fail to capture the detailed dissolution process. {Given the larger amount of water being transported through the PDMS device in the case of groups of droplets, a more precise geometry of the PDMS device would be required to be simulated to have a better prediction. Our results reveal the crucial importance of water transport through the PDMS device in these processes.}

In conclusion, in this work we have employed analytical and numerical tools to estimate the dissolution of water droplets in silicone oil, contained in PDMS-based microfluidic devices. Our work shows the crucial role of the permeability of PDMS to water vapour on the shrinkage of picoliter droplets. Understanding this phenomenon is crucial for microfluidic long-term processes as droplet-based PCR \citep{Prakash2006PCR} or the cultivation of micro-organisms \citep{dewan2012Growth}.\\ 




\section*{Acknowledgements}
The authors acknowledge the financial support from the European Research Council via the Starting Grant \emph{ERC-2015-STG} 678573 and the Advanced Grant \emph{ERC-2016-ADG} 740479. JM and YC acknowledge fruitful discussions with Yaxing Li. The authors specially acknowledge the use of the IBM code developed by Roberto Verzicco and Steven Chong and their assistance.

\bibliographystyle{jfm}

\begin{thebibliography}{55}
\expandafter\ifx\csname natexlab\endcsname\relax\def\natexlab#1{#1}\fi
\def\au#1{#1} \def\ed#1{#1} \def\yr#1{#1}\def\at#1{#1}\def\jt#1{\textit{#1}}
  \def\bt#1{#1}\def\bvol#1{\textbf{#1}} \def\vol#1{#1} \def\pg#1{#1}
  \def\publ#1{#1}\def\arxiv#1{#1}\def\org#1{#1}\def\st#1{\textit{#1}}

\bibitem[Blume {\em et~al.\/}(1991)Blume, Schwering, Mulder \&
  Smolders]{blume1991}
{\sc \au{Blume, I}, \au{Schwering, PJF}, \au{Mulder, MHV} \& \au{Smolders, CA}}
  \yr{1991}  \at{Vapour sorption and permeation properties of poly
  (dimethylsiloxane) films}.  \jt{Journal of Membrane Science}  \bvol{61},
  \pg{85--97}.

\bibitem[Brugarolas {\em et~al.\/}(2013)Brugarolas, Tu \& Lee]{brugarolas2013}
{\sc \au{Brugarolas, Teresa}, \au{Tu, Fuquan} \& \au{Lee, Daeyeon}} \yr{2013}
  \at{Directed assembly of particles using microfluidic droplets and bubbles}.
  \jt{Soft Matter}  \bvol{9}~(38),  \pg{9046--9058}.

\bibitem[Bruning {\em et~al.\/}(2019)Bruning, Costalonga, Snoeijer \&
  Marin]{bruning2019turning}
{\sc \au{Bruning, MA}, \au{Costalonga, M}, \au{Snoeijer, JH} \& \au{Marin, A}}
  \yr{2019}  \at{Turning drops into bubbles: Cavitation by vapor diffusion
  through elastic networks}.  \jt{Physical Review Letters}  \bvol{123}~(21),
  \pg{214501}.

\bibitem[Chong {\em et~al.\/}(2020)Chong, Li, Ng, Verzicco \&
  Lohse]{chong2020convection}
{\sc \au{Chong, Kai~Leong}, \au{Li, Yanshen}, \au{Ng, Chong~Shen},
  \au{Verzicco, Roberto} \& \au{Lohse, Detlef}} \yr{2020}
  \at{Convection-dominated dissolution for single and multiple immersed sessile
  droplets}.  \jt{Journal of Fluid Mechanics}  \bvol{892}.

\bibitem[Cussler(2009)]{cussler2009}
{\sc \au{Cussler, Edward~Lansing}} \yr{2009} {\em Diffusion: mass transfer in
  fluid systems\/}.  \publ{Cambridge university press}.

\bibitem[Dewan {\em et~al.\/}(2012)Dewan, Kim, McLean, Vanapalli \&
  Karim]{dewan2012Growth}
{\sc \au{Dewan, Alim}, \au{Kim, Jihye}, \au{McLean, Rebecca~H}, \au{Vanapalli,
  Siva~A} \& \au{Karim, Muhammad~Nazmul}} \yr{2012}  \at{Growth kinetics of
  microalgae in microfluidic static droplet arrays}.  \jt{Biotechnology and
  bioengineering}  \bvol{109}~(12),  \pg{2987--2996}.

\bibitem[Diehl {\em et~al.\/}(2006)Diehl, Li, He, Kinzler, Vogelstein \&
  Dressman]{diehl2006beaming}
{\sc \au{Diehl, Frank}, \au{Li, Meng}, \au{He, Yiping}, \au{Kinzler,
  Kenneth~W}, \au{Vogelstein, Bert} \& \au{Dressman, Devin}} \yr{2006}
  \at{Beaming: single-molecule pcr on microparticles in water-in-oil
  emulsions}.  \jt{Nature methods}  \bvol{3}~(7),  \pg{551}.

\bibitem[Dollet {\em et~al.\/}(2019)Dollet, Louf, Alonzo, Jensen \&
  Marmottant]{dollet2019drying}
{\sc \au{Dollet, Benjamin}, \au{Louf, Jean-Fran{\c{c}}ois}, \au{Alonzo,
  Mathieu}, \au{Jensen, Kaare~H} \& \au{Marmottant, Philippe}} \yr{2019}
  \at{Drying of channels by evaporation through a permeable medium}.
  \jt{Journal of the Royal Society Interface}  \bvol{16}~(151),  \pg{20180690}.

\bibitem[Duncan \& Needham(2004)]{duncan2004}
{\sc \au{Duncan, P~Brent} \& \au{Needham, David}} \yr{2004}  \at{Test of the
  epstein- plesset model for gas microparticle dissolution in aqueous media:
  Effect of surface tension and gas undersaturation in solution}.
  \jt{Langmuir}  \bvol{20}~(7),  \pg{2567--2578}.

\bibitem[Duncan \& Needham(2006)]{duncan2006}
{\sc \au{Duncan, P~Brent} \& \au{Needham, David}} \yr{2006}  \at{Microdroplet
  dissolution into a second-phase solvent using a micropipet technique: Test of
  the epstein- plesset model for an aniline- water system}.  \jt{Langmuir}
  \bvol{22}~(9),  \pg{4190--4197}.

\bibitem[Epstein \& Plesset(1950)]{epstein1950}
{\sc \au{Epstein, Paul~S} \& \au{Plesset, Milton~S}} \yr{1950}  \at{On the
  stability of gas bubbles in liquid-gas solutions}.  \jt{The Journal of
  Chemical Physics}  \bvol{18}~(11),  \pg{1505--1509}.

\bibitem[Garbay {\em et~al.\/}(1984)Garbay, Grob, Casanovas \&
  Crine]{garbay1984}
{\sc \au{Garbay, H{\^e}l{\`e}ne}, \au{Grob, Robert}, \au{Casanovas, Joseph} \&
  \au{Crine, Jean-Pierre}} \yr{1984} Measurements and influence of water
  content in silicone oil.  \bt{In {\em Electrical Insulation, 1984 IEEE
  International Conference on\/}},  \pg{pp. 297--300}. IEEE.

\bibitem[Hilder \& van~den Tempe(1971)]{hilder1971}
{\sc \au{Hilder, MH} \& \au{van~den Tempe, M}} \yr{1971}  \at{Diffusivity of
  water in groundnut oil and paraffi oil}.  \jt{Journal of Chemical Technology
  and Biotechnology}  \bvol{21}~(6),  \pg{176--178}.

\bibitem[Jain \& Verma(2011)]{jain2011}
{\sc \au{Jain, Archana} \& \au{Verma, Krishna~K}} \yr{2011}  \at{Recent
  advances in applications of single-drop microextraction: a review}.
  \jt{Analytica Chimica Acta}  \bvol{706}~(1),  \pg{37--65}.

\bibitem[de~Jong {\em et~al.\/}(2006)de~Jong, Lammertink \&
  Wessling]{deJong:2006jt}
{\sc \au{de~Jong, J}, \au{Lammertink, R G~H} \& \au{Wessling, M}} \yr{2006}
  \at{{Membranes and microfluidics: a review}}.  \jt{Lab Chip}  \bvol{6}~(9),
  \pg{1125--15}.

\bibitem[Laghezza {\em et~al.\/}(2016)Laghezza, Dietrich, Yeomans,
  Ledesma-Aguilar, Kooij, Zandvliet \& Lohse]{Laghezza:2016zr}
{\sc \au{Laghezza, Gianluca}, \au{Dietrich, Erik}, \au{Yeomans, Julia~M.},
  \au{Ledesma-Aguilar, Rodrigo}, \au{Kooij, E.~Stefan}, \au{Zandvliet, Harold
  J.~W.} \& \au{Lohse, Detlef}} \yr{2016}  \at{Collective and convective
  effects compete in patterns of dissolving surface droplets}.  \jt{Soft
  Matter}  \bvol{12}~(26),  \pg{5787--5796}.

\bibitem[Lauga \& Brenner(2004)]{Lauga:2004qy}
{\sc \au{Lauga, Eric} \& \au{Brenner, Michael~P.}} \yr{2004}
  \at{Evaporation-driven assembly of colloidal particles}.  \jt{Physical Review
  Letters}  \bvol{93}~(23),  \pg{238301--}.

\bibitem[Leng {\em et~al.\/}(2006)Leng, Lonetti, Tabeling, Joanicot \&
  Ajdari]{LengPRL2006microevaporators}
{\sc \au{Leng, Jacques}, \au{Lonetti, Barbara}, \au{Tabeling, Patrick},
  \au{Joanicot, Mathieu} \& \au{Ajdari, Armand}} \yr{2006}
  \at{Microevaporators for kinetic exploration of phase diagrams}.
  \jt{Physical review letters}  \bvol{96}~(8),  \pg{084503}.

\bibitem[Lohse \& Zhang(2015)]{Lohse2015RMP}
{\sc \au{Lohse, Detlef} \& \au{Zhang, Xuehua}} \yr{2015}  \at{Surface
  nanobubbles and nanodroplets}.  \jt{Reviews of Modern Physics}
  \bvol{87}~(3),  \pg{981}.

\bibitem[Manoharan {\em et~al.\/}(2003)Manoharan, Elsesser \&
  Pine]{manoharan2003}
{\sc \au{Manoharan, Vinothan~N}, \au{Elsesser, Mark~T} \& \au{Pine, David~J}}
  \yr{2003}  \at{Dense packing and symmetry in small clusters of microspheres}.
   \jt{Science}  \bvol{301}~(5632),  \pg{483--487}.

\bibitem[Margulies {\em et~al.\/}(2005)Margulies, Egholm, Altman, Attiya,
  Bader, Bemben, Berka, Braverman, Chen, Chen {\em
  et~al.\/}]{margulies2005genome}
{\sc \au{Margulies, Marcel}, \au{Egholm, Michael}, \au{Altman, William~E},
  \au{Attiya, Said}, \au{Bader, Joel~S}, \au{Bemben, Lisa~A}, \au{Berka, Jan},
  \au{Braverman, Michael~S}, \au{Chen, Yi-Ju}, \au{Chen, Zhoutao} \&
  \au{others}} \yr{2005}  \at{Genome sequencing in microfabricated high-density
  picolitre reactors}.  \jt{Nature}  \bvol{437}~(7057),  \pg{376}.

\bibitem[Michelin {\em et~al.\/}(2018)Michelin, Gu{\'e}rin \&
  Lauga]{michelin2018collective}
{\sc \au{Michelin, S{\'e}bastien}, \au{Gu{\'e}rin, Etienne} \& \au{Lauga,
  Eric}} \yr{2018}  \at{Collective dissolution of microbubbles}.  \jt{Physical
  Review Fluids}  \bvol{3}~(4),  \pg{043601}.

\bibitem[Noblin {\em et~al.\/}(2008)Noblin, Mahadevan, Coomaraswamy, Weitz,
  Holbrook \& Zwieniecki]{noblin2008optimal}
{\sc \au{Noblin, X}, \au{Mahadevan, Lakshminarayanan}, \au{Coomaraswamy, IA},
  \au{Weitz, David~A}, \au{Holbrook, Noel~Michele} \& \au{Zwieniecki,
  Maciej~A}} \yr{2008}  \at{Optimal vein density in artificial and real
  leaves}.  \jt{Proceedings of the National Academy of Sciences}
  \bvol{105}~(27),  \pg{9140--9144}.

\bibitem[Prakash {\em et~al.\/}(2006)Prakash, Adamia, Sieben, Pilarski,
  Pilarski \& Backhouse]{Prakash2006PCR}
{\sc \au{Prakash, A~Ranjit}, \au{Adamia, S}, \au{Sieben, V}, \au{Pilarski, P},
  \au{Pilarski, LM} \& \au{Backhouse, CJ}} \yr{2006}  \at{Small volume pcr in
  pdms biochips with integrated fluid control and vapour barrier}.  \jt{Sensors
  and Actuators B: Chemical}  \bvol{113}~(1),  \pg{398--409}.

\bibitem[Randall \& Doyle(2005)]{DoylePNAS2005}
{\sc \au{Randall, Greg~C} \& \au{Doyle, Patrick~S}} \yr{2005}
  \at{Permeation-driven flow in poly (dimethylsiloxane) microfluidic devices}.
  \jt{Proceedings of the National Academy of Sciences}  \bvol{102}~(31),
  \pg{10813--10818}.

\bibitem[R\'e(1998)]{spraydrying1998}
{\sc \au{R\'e, M.~I.}} \yr{1998}  \at{Microencapsulation by spray drying}.
  \jt{Drying Technology}  \bvol{16}~(6),  \pg{1195--1236}.

\bibitem[Rezaee {\em et~al.\/}(2006)Rezaee, Assadi, Hosseini, Aghaee, Ahmadi \&
  Berijani]{rezaee2006}
{\sc \au{Rezaee, Mohammad}, \au{Assadi, Yaghoub}, \au{Hosseini,
  Mohammad-Reza~Milani}, \au{Aghaee, Elham}, \au{Ahmadi, Fardin} \&
  \au{Berijani, Sana}} \yr{2006}  \at{Determination of organic compounds in
  water using dispersive liquid--liquid microextraction}.  \jt{Journal of
  Chromatography A}  \bvol{1116}~(1-2),  \pg{1--9}.

\bibitem[Rezaee {\em et~al.\/}(2010)Rezaee, Yamini \& Faraji]{rezaee2010}
{\sc \au{Rezaee, Mohammad}, \au{Yamini, Yadollah} \& \au{Faraji, Mohammad}}
  \yr{2010}  \at{Evolution of dispersive liquid--liquid microextraction
  method}.  \jt{Journal of Chromatography A}  \bvol{1217}~(16),
  \pg{2342--2357}.

\bibitem[Rivero-Rodriguez \& Scheid(2019)]{Rivero2019}
{\sc \au{Rivero-Rodriguez, Javier} \& \au{Scheid, Benoit}} \yr{2019}  \at{Mass
  transfer around bubbles flowing in cylindrical microchannels}.  \jt{Journal
  of Fluid Mechanics}  \bvol{869},  \pg{110–142}.

\bibitem[Robb(1968)]{Rubber1968}
{\sc \au{Robb, WL}} \yr{1968}  \at{Thin silicone membranes-their permeation
  properties and some applications}.  \jt{Annals of the New York Academy of
  Sciences}  \bvol{146}~(1),  \pg{119--137}.

\bibitem[Rydberg(2004)]{rydberg2004solvent}
{\sc \au{Rydberg, Jan}} \yr{2004} {\em Solvent extraction principles and
  practice, revised and expanded\/}.  \publ{CRC Press, New York, USA}.

\bibitem[Shah {\em et~al.\/}(2008)Shah, Shum, Rowat, Lee, Agresti, Utada, Chu,
  Kim, Fernandez-Nieves, Martinez {\em et~al.\/}]{shah2008designer}
{\sc \au{Shah, Rhutesh~K}, \au{Shum, Ho~Cheung}, \au{Rowat, Amy~C}, \au{Lee,
  Daeyeon}, \au{Agresti, Jeremy~J}, \au{Utada, Andrew~S}, \au{Chu, Liang-Yin},
  \au{Kim, Jin-Woong}, \au{Fernandez-Nieves, Alberto}, \au{Martinez, Carlos~J}
  \& \au{others}} \yr{2008}  \at{Designer emulsions using microfluidics}.
  \jt{Materials Today}  \bvol{11}~(4),  \pg{18--27}.

\bibitem[Shim {\em et~al.\/}(2014)Shim, Wan, Hilgenfeldt, Panchal \&
  Stone]{shim2014dissolution}
{\sc \au{Shim, Suin}, \au{Wan, Jiandi}, \au{Hilgenfeldt, Sascha}, \au{Panchal,
  Prathamesh~D} \& \au{Stone, Howard~A}} \yr{2014}  \at{Dissolution without
  disappearing: Multicomponent gas exchange for co 2 bubbles in a microfluidic
  channel}.  \jt{Lab on a Chip}  \bvol{14}~(14),  \pg{2428--2436}.

\bibitem[Song {\em et~al.\/}(2003)Song, Tice \& Ismagilov]{Song:2003bg}
{\sc \au{Song, Helen}, \au{Tice, Joshua~D.} \& \au{Ismagilov, Rustem~F.}}
  \yr{2003}  \at{{A Microfluidic System for Controlling Reaction Networks in
  Time}}.  \jt{Angew. Chem.}  \bvol{115}.

\bibitem[Stone {\em et~al.\/}(2004)Stone, Stroock \&
  Ajdari]{stone2004engineering}
{\sc \au{Stone, Howard~A}, \au{Stroock, Abraham~D} \& \au{Ajdari, Armand}}
  \yr{2004}  \at{Engineering flows in small devices: microfluidics toward a
  lab-on-a-chip}.  \jt{Annu. Rev. Fluid Mech.}  \bvol{36},  \pg{381--411}.

\bibitem[Takeuchi {\em et~al.\/}(2005)Takeuchi, Garstecki, Weibel \&
  Whitesides]{Takeuchi:2005dl}
{\sc \au{Takeuchi, S}, \au{Garstecki, P}, \au{Weibel, D~B} \& \au{Whitesides,
  G~M}} \yr{2005}  \at{{An Axisymmetric Flow-Focusing Microfluidic Device}}.
  \jt{Adv. Mater.}  \bvol{17}~(8),  \pg{1067--1072}.

\bibitem[This(2002)]{this2002molecular}
{\sc \au{This, Herv{\'e}}} \yr{2002}  \at{Molecular gastronomy}.
  \jt{Angewandte Chemie International Edition}  \bvol{41}~(1),  \pg{83--88}.

\bibitem[This(2005)]{this2005molecular}
{\sc \au{This, Herve}} \yr{2005}  \at{Molecular gastronomy}.  \jt{Nature
  Materials}  \bvol{4}~(1),  \pg{5}.

\bibitem[Vega-Mart\'inez {\em et~al.\/}(2020)Vega-Mart\'inez,
  Rodr\'iguez-Rodr\'iguez \& van~der Meer]{Patri2020}
{\sc \au{Vega-Mart\'inez, Patricia}, \au{Rodr\'iguez-Rodr\'iguez, Javier} \&
  \au{van~der Meer, Devaraj}} \yr{2020}  \at{Growth of a bubble cloud in
  co2-saturated water under microgravity}.  \jt{Soft Matter}  \bvol{16},
  \pg{4728--4738}.

\bibitem[Velev {\em et~al.\/}(2000)Velev, Lenhoff \& Kaler]{velev2000}
{\sc \au{Velev, Orlin~D}, \au{Lenhoff, Abraham~M} \& \au{Kaler, Eric~W}}
  \yr{2000}  \at{A class of microstructured particles through colloidal
  crystallization}.  \jt{Science}  \bvol{287}~(5461),  \pg{2240--2243}.

\bibitem[Verneuil {\em et~al.\/}(2004)Verneuil, Buguin \&
  Silberzan]{Verneuil2004permeation}
{\sc \au{Verneuil, E}, \au{Buguin, A} \& \au{Silberzan, P}} \yr{2004}
  \at{Permeation-induced flows: Consequences for silicone-based microfluidics}.
   \jt{EPL (Europhysics Letters)}  \bvol{68}~(3),  \pg{412}.

\bibitem[Vincent {\em et~al.\/}(2012)Vincent, Marmottant, Quinto-Su \&
  Ohl]{vincent2012birth}
{\sc \au{Vincent, Olivier}, \au{Marmottant, Philippe}, \au{Quinto-Su, Pedro~A}
  \& \au{Ohl, Claus-Dieter}} \yr{2012}  \at{Birth and growth of cavitation
  bubbles within water under tension confined in a simple synthetic tree}.
  \jt{Physical Review Letters}  \bvol{108}~(18),  \pg{184502}.

\bibitem[Volk {\em et~al.\/}(2015)Volk, Rossi, K\"ahler, Hilgenfeldt \&
  Marin]{volk:2015}
{\sc \au{Volk, Andreas}, \au{Rossi, Massimiliano}, \au{K\"ahler, Christian~J.},
  \au{Hilgenfeldt, Sascha} \& \au{Marin, Alvaro}} \yr{2015}  \at{Growth control
  of sessile microbubbles in pdms devices}.  \jt{Lab on a Chip}  \bvol{15},
  \pg{4607--4615}.

\bibitem[Wang {\em et~al.\/}(2017)Wang, Jin, Gong, Li, Wu, Zhang, Zhou, Shui,
  Eijkel \& van~den Berg]{wang2017}
{\sc \au{Wang, Juan}, \au{Jin, Mingliang}, \au{Gong, Yingxin}, \au{Li, Hao},
  \au{Wu, Sujuan}, \au{Zhang, Zhang}, \au{Zhou, Guofu}, \au{Shui, Lingling},
  \au{Eijkel, Jan~CT} \& \au{van~den Berg, Albert}} \yr{2017}  \at{Continuous
  fabrication of microcapsules with controllable metal covered nanoparticle
  arrays using droplet microfluidics for localized surface plasmon resonance}.
  \jt{Lab on a Chip}  \bvol{17}~(11),  \pg{1970--1979}.

\bibitem[Wang {\em et~al.\/}(2018)Wang, Mbah, Przybilla, Zubiri, Spiecker,
  Engel \& Vogel]{Vogel2018magic}
{\sc \au{Wang, Junwei}, \au{Mbah, Chrameh~Fru}, \au{Przybilla, Thomas},
  \au{Zubiri, Benjamin~Apeleo}, \au{Spiecker, Erdmann}, \au{Engel, Michael} \&
  \au{Vogel, Nicolas}} \yr{2018}  \at{Magic number colloidal clusters as
  minimum free energy structures}.  \jt{Nature communications}  \bvol{9}~(1),
  \pg{1--10}.

\bibitem[Watson \& Baron(1996)]{watson1996}
{\sc \au{Watson, JM} \& \au{Baron, MG}} \yr{1996}  \at{The behaviour of water
  in poly (dimethylsiloxane)}.  \jt{Journal of Membrane Science}
  \bvol{110}~(1),  \pg{47--57}.

\bibitem[Wheeler \& Stroock(2008)]{wheeler2008transpiration}
{\sc \au{Wheeler, Tobias~D} \& \au{Stroock, Abraham~D}} \yr{2008}  \at{The
  transpiration of water at negative pressures in a synthetic tree}.
  \jt{Nature}  \bvol{455}~(7210),  \pg{208--212}.

\bibitem[Xia \& Whitesides(1998{\natexlab{{\em a\/}}})]{xia1998soft}
{\sc \au{Xia, Younan} \& \au{Whitesides, George~M}} \yr{1998{\natexlab{{\em
  a\/}}}}  \at{Soft lithography}.  \jt{Annual review of materials science}
  \bvol{28}~(1),  \pg{153--184}.

\bibitem[Xia \& Whitesides(1998{\natexlab{{\em b\/}}})]{xia1998}
{\sc \au{Xia, Younan} \& \au{Whitesides, George~M}} \yr{1998{\natexlab{{\em
  b\/}}}}  \at{Soft lithography}.  \jt{Annual Review of Materials Science}
  \bvol{28}~(1),  \pg{153--184}.

\bibitem[Yi {\em et~al.\/}(2003)Yi, Thorsen, Manoharan, Hwang, Jeon, Pine,
  Quake \& Yang]{yi2003}
{\sc \au{Yi, G-R}, \au{Thorsen, Todd}, \au{Manoharan, Vinothan~N}, \au{Hwang,
  M-J}, \au{Jeon, S-J}, \au{Pine, David~J}, \au{Quake, Stephan~R} \& \au{Yang,
  S-M}} \yr{2003}  \at{Generation of uniform colloidal assemblies in soft
  microfluidic devices}.  \jt{Advanced Materials}  \bvol{15}~(15),
  \pg{1300--1304}.

\bibitem[Yu {\em et~al.\/}(2012)Yu, Wang, Oberthuer, Meyer, Perbandt, Duan \&
  Kang]{yu2012}
{\sc \au{Yu, Yong}, \au{Wang, Xuan}, \au{Oberthuer, Dominik}, \au{Meyer, Arne},
  \au{Perbandt, Markus}, \au{Duan, Li} \& \au{Kang, Qi}} \yr{2012}  \at{Design
  and application of a microfluidic device for protein crystallization using an
  evaporation-based crystallization technique}.  \jt{Journal of Applied
  Crystallography}  \bvol{45}~(1),  \pg{53--60}.

\bibitem[Zhang {\em et~al.\/}(2012)Zhang, Coulston, Jones, Geng, Scherman \&
  Abell]{zhang2012one}
{\sc \au{Zhang, Jing}, \au{Coulston, Roger~J}, \au{Jones, Samuel~T}, \au{Geng,
  Jin}, \au{Scherman, Oren~A} \& \au{Abell, Chris}} \yr{2012}  \at{One-step
  fabrication of supramolecular microcapsules from microfluidic droplets}.
  \jt{Science}  \bvol{335}~(6069),  \pg{690--694}.

\bibitem[Zheng {\em et~al.\/}(2003)Zheng, Roach \& Ismagilov]{zheng2003}
{\sc \au{Zheng, Bo}, \au{Roach, L~Spencer} \& \au{Ismagilov, Rustem~F}}
  \yr{2003}  \at{Screening of protein crystallization conditions on a
  microfluidic chip using nanoliter-size droplets}.  \jt{Journal of the
  American Chemical Society}  \bvol{125}~(37),  \pg{11170--11171}.

\bibitem[Zhu \& Wang(2017)]{zhu2017}
{\sc \au{Zhu, Pingan} \& \au{Wang, Liqiu}} \yr{2017}  \at{Passive and active
  droplet generation with microfluidics: a review}.  \jt{Lab on a Chip}
  \bvol{17}~(1),  \pg{34--75}.

\bibitem[Zhu {\em et~al.\/}(2018)Zhu, Verzicco, Zhang \& Lohse]{Zhu:2018}
{\sc \au{Zhu, Xiaojue}, \au{Verzicco, Roberto}, \au{Zhang, Xuehua} \&
  \au{Lohse, Detlef}} \yr{2018}  \at{Diffusive interaction of multiple surface
  nanobubbles: shrinkage{,} growth{,} and coarsening}.  \jt{Soft Matter}
  \bvol{14},  \pg{2006--2014}.

\end{thebibliography}

\end{document}